\documentclass[twocolumn,10pt]{IEEEtran}
\usepackage{amsmath,amssymb,epsfig,graphicx}
\usepackage{theorem}
\newcommand{\be}{\begin{equation}}
\newcommand{\ee}{\end{equation}}

\title{Effects of variations of load distribution on network performance}
\author{
David Arrowsmith$^{\dag},$\thanks{$^{\dag}$ Mathematics Research
Centre, Queen Mary, University of London, London E1 4NS, U.K.}
Mario di Bernardo$^{*\circ},$\thanks{$^{*}$Department of
Engineering Mathematics, University of Bristol, Bristol, U.K.}
Francesco Sorrentino$^{\circ \ddag}$\thanks{$^{°\circ}$Department
of Systems and Computer Science, University of Naples Federico II,
Naples, Italy} \thanks{$^{°\ddag}$Corresponding author. Email:
{\small fsorrent@unina.it}}}

\date{}

\begin{document}
%
\maketitle
\begin{abstract}
This paper is concerned with the characterization of the
relationship between topology and traffic dynamics. We use a model
of network generation that allows the transition from random to
scale free networks.  Specifically, we consider three different
topological types of network: random, scale-free with $\gamma=3$,
scale-free with $\gamma=2$. By using a novel LRD traffic
generator, we observe best performance, in terms of transmission
rates and delivered packets, in the case of random networks. We
show that, even if scale-free networks are characterized by
shorter characteristic-path-length (the lower the exponent, the
lower the path-length), they show worst performances in terms of
communication. We conjecture this could be explained in terms of
changes in the load distribution, defined here as the number of
shortest paths going through a given vertex. In fact, that
distribution is characterized by (i) a decreasing mean (ii) an
increasing standard deviation, as the networks becomes scale-free
(especially scale-free networks with low exponents). The use of a
degree-independent server also discriminates against a scale-free
structure. As a result, since the model is uncontrolled, most
packets will go through the same vertices, favoring the onset of
congestion.
\end{abstract}
\section{Introduction}
\label{sec:intro}

Much research effort has been spent recently in understanding the
relationship between network topological features and
communication performances. In \cite{korea}, the problem of
relating the degree distribution (i.e. the distribution of the
number of incident links at a given node) to the load distribution
(that is the number of shortest paths passing through a given
vertex) in a given network is discussed. The load parameter is
shown to be useful to give a statistical measure of the
probability that a generic packet, travelling in the network, will
pass through a given vertex. Nevertheless, it is not taken into
account that, in real world applications, packets are stored in
routers' queues while going from origin to destination, causing
time delays in the communication.

  As a first approximation, it would be natural to make the most
general hypothesis about the structure of the underlying network,
that is, to think of it as a random graph. Unfortunately, real
networks show statistical properties that are far from being
completely random. The most important difference is that they have
typically power law degree distributions with exponents between 2
and 3 \cite{Am:Sc00}. For that reason here we have considered
three different topologies, in the order: random, scale-free with
$\gamma=3$, scale-free with $\gamma=2$.

In this paper, we will consider a packet transport model that has
been widely studied in the literature (see \cite{Oh:Sa},
\cite{So:Va}, \cite{ARROW04} for further details), in order to
compare the main indicators of the network performance,
specifically the delivery time and the number of delivered packets
(or throughput), as the underlying topology is varied.

\section{From Random to Scale-Free Networks}
\label{sec:application}

As a first general approximation we can think of the underlying
network to be represented by a random graph (cf. the well-known
model by Erdos and Renyi presented in \cite{Er:Re}). A network
with $N$ vertices and $M$ edges is built as follows: we select
with uniform probability two of the $N$ possible vertices and link
them unless they are either already connected or self-links are
generated. We repeat this iteration $M$ times.

While the ER graph is pioneering, various properties of this model
are not in accordance with those of complex networks recently
discovered in the real world. For example, the distribution of the
number of edges incident on each vertex, called the {\it degree
distribution}, is Poissonian for the ER graph, while it follows a
power law decay with increasing degree for many real world
networks, called scale free (SF) networks \cite{Ba:Al99},
\cite{Fa:Fa99}, \cite{Am:Sc00}, \cite{Do:Me02}. That is real
networks differ deeply from ER networks since they are
characterized by a power-law degree distribution.

In order to cause the transition from random to scale-free network
we use the static model recently introduced in \cite{korea}.
Vertices are indexed by an integer $i$, for $(i=1....,N)$, and
assigned a {\it weight} or {\it fitness} $p_i=i^{-\alpha}$ where
$\alpha$ is a parameter between 0 and 1. Two different vertices
are selected with probabilities equal to the normalized weights,
$p_i/\sum_k{p_k}$ and $p_j/\sum_k{p_k}$ respectively and an edge
is added between them unless one exists already. This process is
repeated until $M$ edges are made in the system leading to the
mean degree $\langle k \rangle = 2M/N $. This results in the
expected degree at vertex $i$ scaling as $k_i \sim
(\frac{N}{i})^\alpha$ \cite{korea}. We then have the degree
distribution, i.e. the probability of a vertex being of degree
$k$, given by $P(k)\sim k^{-\gamma}$ with $ \gamma= 1+
\frac{1}{\alpha}$. Thus, by varying $\alpha$, we can  obtain the
exponent $\gamma$ in the range, $2<\gamma<\infty$. Moreover the ER
graph is generated by taking $\alpha=0$.

It is worth noting that the static model described here, can be
considered as an extension of the standard ER model for generating
\emph{random-scale free networks}, i.e. networks with prescribed
degree distribution, but completely random with respect to all the
other features.

\section{Load distribution in Networks}

One of the main parameters of vertex centrality is the
\emph{betweenness centrality} defined as the number of shortest
paths between pairs of nodes crossing a given vertex
\cite{FreeBOOK}. With this index as a starting point, Goh {\it et
al.} \cite{korea} \cite{korea3}, defined the load at each vertex
$v$, say $l(v)$, as the number of packets passing through it,
under the assumption that every node sends a packet to every other
node in the network and that packets move in parallel from origin
to destination through the geodesic, i.e. the shortest path
between them. This implies that for each shortest path between a
given couple of vertices, there is a packet passing along it; in
the case that packets encounter a branching point at which there
is more than one shortest path toward the destination, they would
be divided by the number of branches at the branching point. Thus,
in \cite{korea} the load at each vertex, is defined as:
\begin{equation}
\label{eq:load1} l(v)=\sum_{s\neq
t}\frac{\varsigma_{st}(v)}{\varsigma_{st}}
\end{equation}
where $\varsigma_{st}(v)$ is the number of shortest paths going
from $s$ to $t$ through $v$ and $\varsigma_{st}$ is the total
number of shortest paths between $s$ and $t$.

Moreover, we measure here another parameter to characterize
further the uniformity of the load distribution: the load standard
deviation $\sigma(l)$  over the vertices of a given network. Here,
to make it insensitive to the average values of $l$, we will
evaluate $\tilde{\sigma}(l)$, the standard deviation of the load
appropriately normalized with respect to its mean.
Specifically, a high variance of that distribution should indicate
an unfair use of the network, and could therefore indicate a
possible cause for congestion. Thus the load standard deviation
gives a measure of how the network topology can lead itself to a
fair exploitation of the network vertices. Note that from a
communications point of view it could be
desirable to minimize $\tilde{\sigma}(l)$ 
in order to make the exploitation of the network resources as fair
as possible.

We have computed the load distribution by varying the $\alpha$
parameter. In Fig.1  the average load is reported while in Fig.2
we depict its standard deviation. Note that for $\alpha=0$, the
resulting topology is the standard ER network; for $\alpha=0.5$,
it is a scale-free with $\gamma=3$; for $\alpha=1$ it is a
scale-free with $\gamma=2$.

As the network transition from random to scale-free occurs, we
observe that the average load decreases, since the presence of
hubs results in a shortening of the mean distances between
vertices. Nevertheless this happens at the expense of the fairness
of the network resources exploitation. Evidence of this is shown
in Fig.2, where the load standard deviation increases with
$\alpha$. This indicates that relatively few vertices are drawing
most of the network load.

It needs to be stressed that the load behavior fails in describing
real communication on networks, since, ideally, packets are
supposed to travel from sources to destinations directly, without
having to be stored along their pathway in the nodes queue. That
is equivalent to assuming that node queues have an infinite
transmission rate.

Nevertheless we will show here that the network communication
behavior is actually affected by the load distribution, since the
load parameter gives the probability for a generic packet to be
forwarded to a given node. In order to better characterise how the
load distribution affects the network performance, we need first
to choose an appropriate traffic generation model.

\begin{figure}[tb]
\begin{center}
\epsfig{width=0.50\textwidth, file=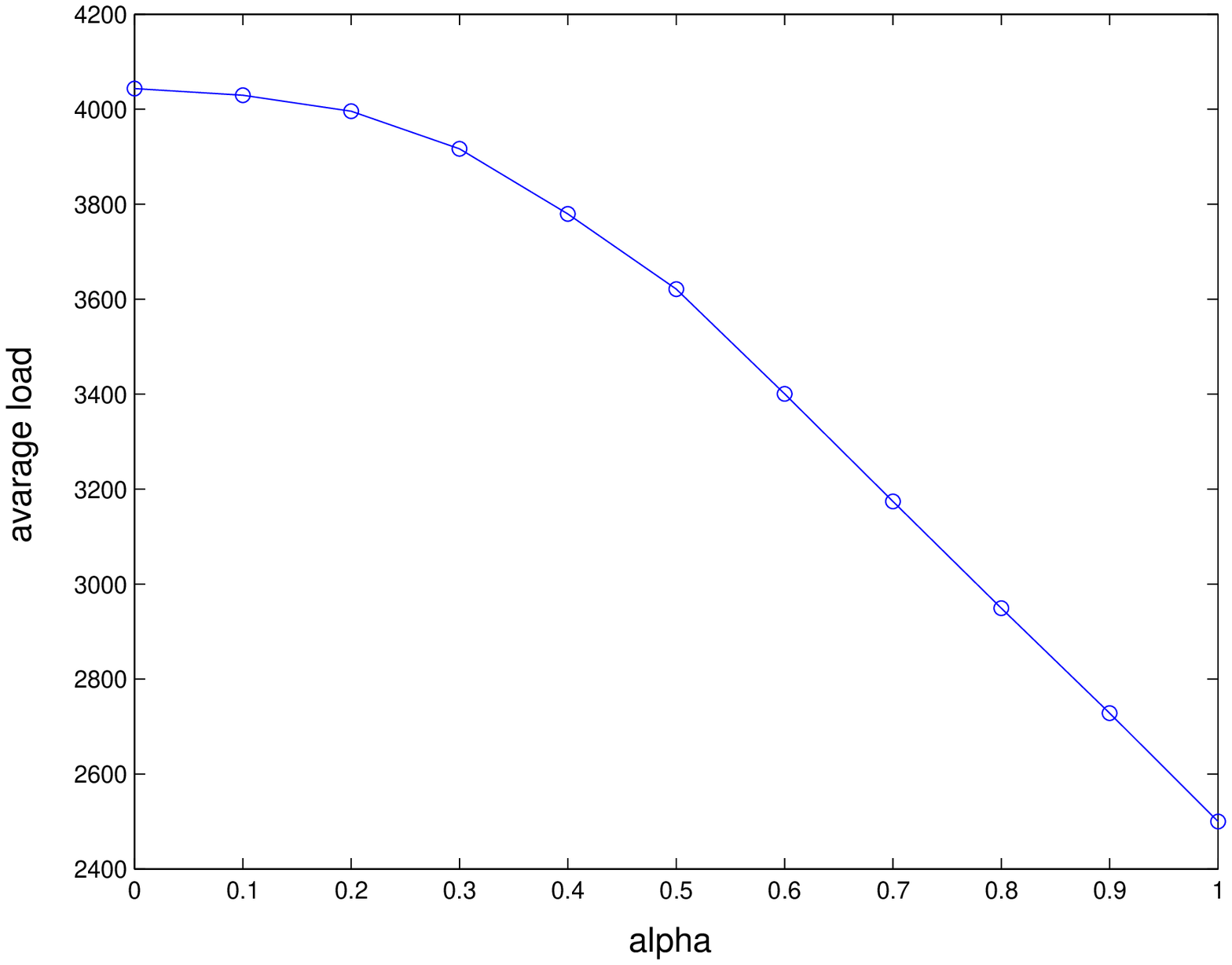}
\end{center}
\caption{\label{fig:SF1} \small The average load at the vertices
is given with respect to the parameter  $\alpha$, while keeping
the number of vertices fixed (at 500) and of the edge degree (3
per node).}
\end{figure}

\begin{figure}[tb]
\begin{center}
\epsfig{width=0.50\textwidth, file=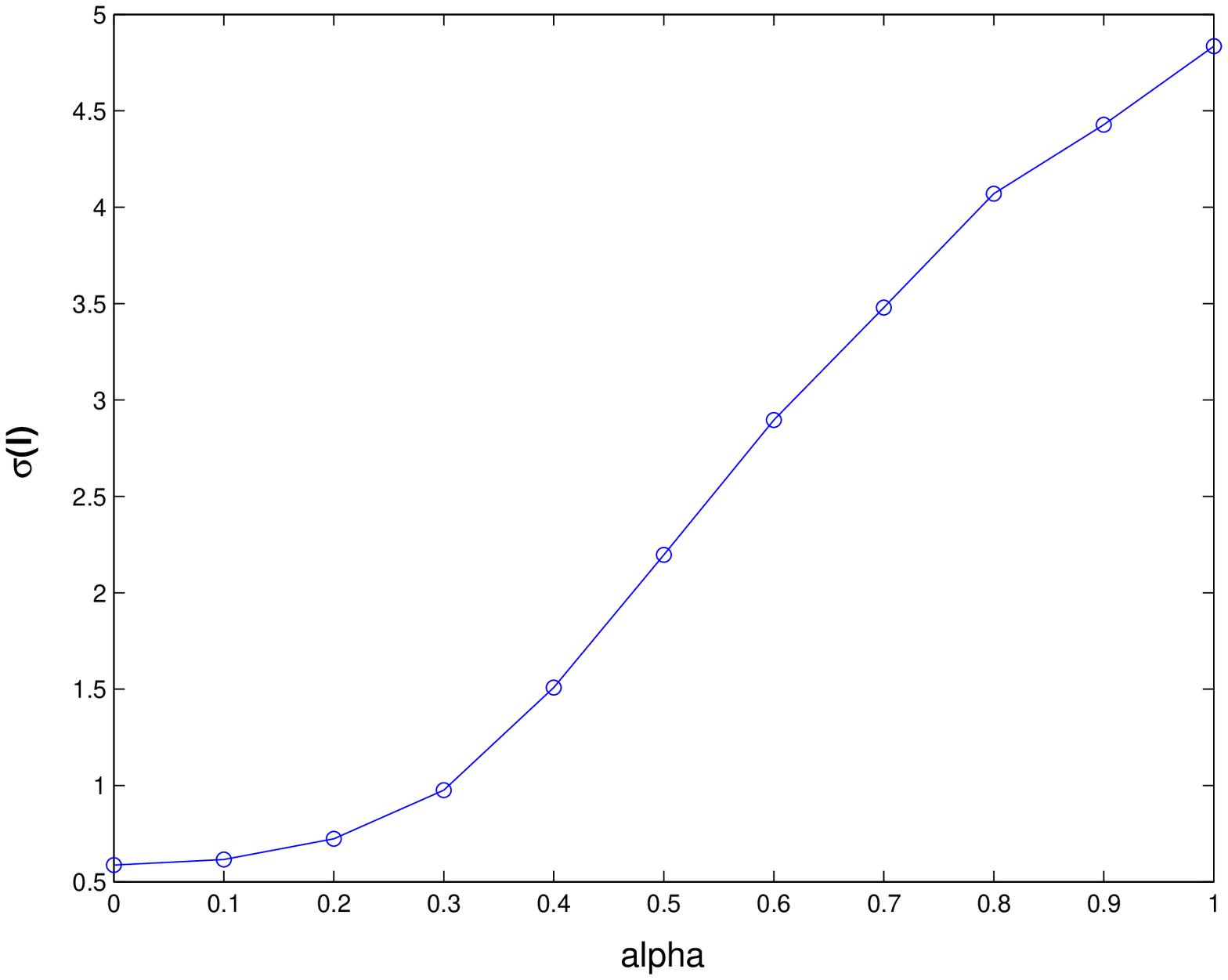}
\end{center}
\caption{\label{fig:SF2} \small Normalized STD of vertex load
$\tilde{\sigma}(l)$ is reported as varying the $\alpha$ parameter,
under the same hypothesis of figure 2.}
\end{figure}

\section{Model of Network Data Traffic}

We use the family of {\it Erramilli} interval maps as the
generator for each LRD traffic source, (Erramilli {\it et al.},
1994),\cite{Erramilli} within the network.  The maps are given by
$f=f_{(m_1,m_2,d)}:I\rightarrow I$, $I=[0,1]$, where:

\be \label{ESPmap} f(x)=\begin{cases}
x+(1-d)\left({x/d}\right)^{m_1}, & x\in [0,d],\cr
x-d\left((1-x)/(1-d)\right)^{m_2}, & x\in (d,1],
\end{cases}
\ee

\noindent where $d\in (0,1)$.  The map $f$ is iterated to produce
an {\it orbit}, or sequence, of real numbers $x_n\in [0,1]$ which
is then converted into a binary {\it Off-On} sequence where the
$n$-th value is  {\it'Off'} if $x_n\in [0,d]$, and {\it 'On'} if
$x_n\in (d,1]$. If the map is in the {\it'On'} state, each
iteration of the map represents a packet generated. The parameters
$m_1,m_2\in [3/2,2]$ induce {\it map intermittency}. When
$m_1=m_2=1.5$ we have {\it short range dependent} binary output
and this becomes fully {\it long range dependent} binary output
for ${\rm Max}\{m_1,m_2\}=2.0$.

\begin{figure}[t]
\begin{center}
\epsfig{width=0.50\textwidth, file=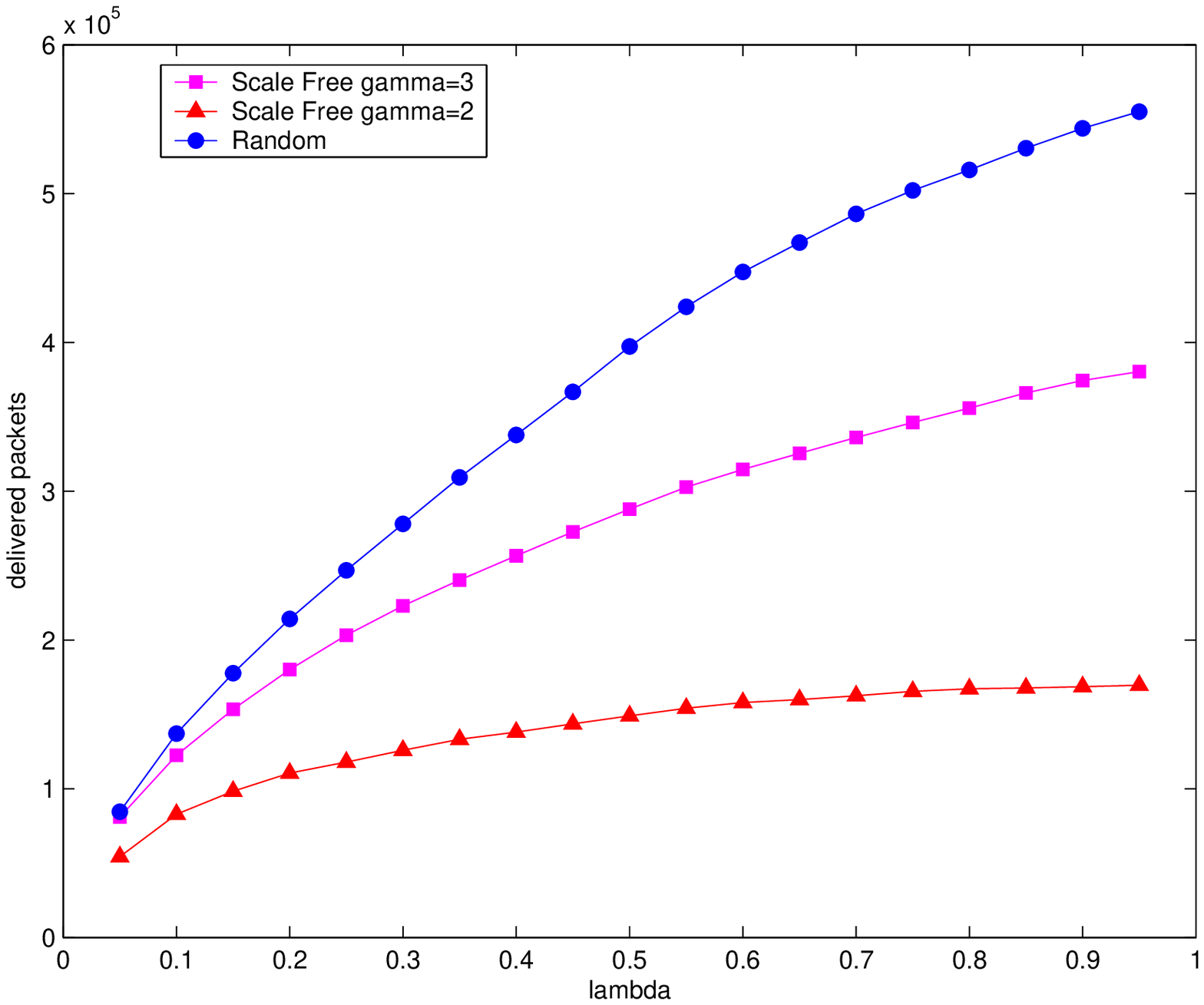}
\end{center}
\caption{\label{fig:SF1} \small Number of delivered packets versus
the generation rate, $\lambda$. Three different networks, random,
scale-free with $\gamma=3$, scale-free with $\gamma=2$, have been
compared, while keeping a fixed number of vertices (500) and the
edge degree (3 per node).}
\end{figure}

The network involves two types of nodes: hosts and routers. The
first are nodes that can generate and receive messages and the
second can only store and forward messages. The density of hosts
$\rho \in [0,1]$ is the ratio between the number of hosts and the
total number of nodes in the network (in this paper we take $\rho
= 0.16$). Hosts are randomly distributed throughout the network.

A routing algorithm is needed to model the dynamic aspects of the
network. Packets are created at hosts and sent through the lattice
one step at a time until they reach their destination host.

The routing algorithm operates as follows:

(1) First a host creates a packet following a distribution defined
by a chaotic map (LRD), as described above. If a packet is
generated it is put at the end of the queue for that host. This is
repeated for each host in the lattice.

(2) Packets at the head of each queue are picked up and sent to a
neighboring node selected according to the following rules. (a) A
neighbor closest to the destination node is selected. (b) If more
than one neighbor is at the minimum distance from the destination,
the link through which the smallest number of packets have been
forwarded is selected. (c) If more than one of these links shares
the same minimum number of packets forwarded, then a random
selection is made.

This process is repeated for each node in the lattice. The whole
procedure of packet generation and movement represents one time
step of the simulation.

\begin{figure}[t]
\begin{center}
\epsfig{width=0.50\textwidth, file=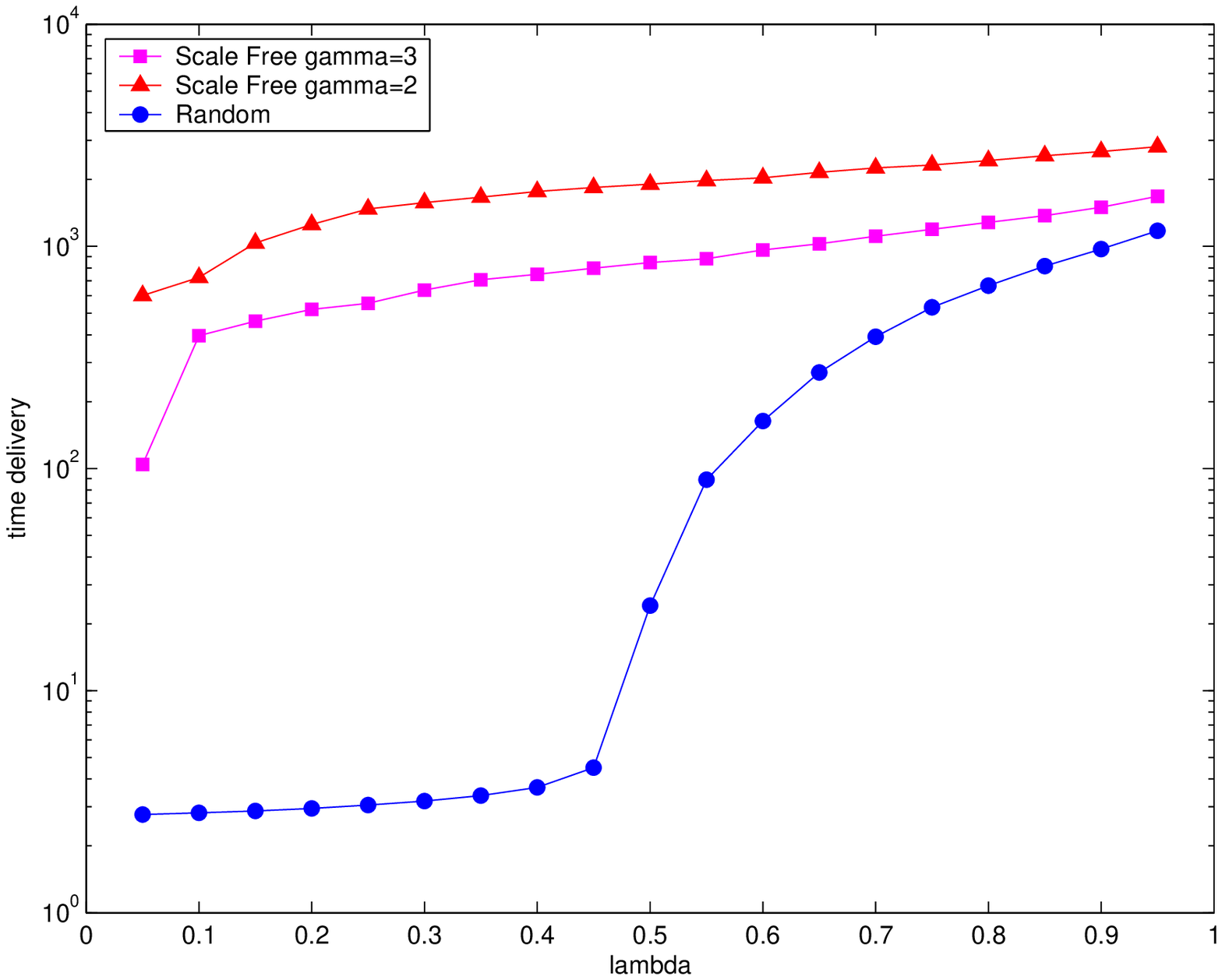}
\end{center}
\caption{\label{fig:SF2} \small Delivery times versus the
generation rate, $\lambda$. Three different networks, random,
scale-free with $\gamma=3$, scale-free with $\gamma=2$, have been
compared, while keeping fixed the number of vertices (500) and the
edges (3 per node)}
\end{figure}

\section{Effects on network performance of varying the underlying topology}

  We have compared three different topologies: random, scale-free
with $\gamma=3$, scale-free with $\gamma=2$ in order to evaluate
the effects of the underlying topology on the network
performances.

The networks we consider have different degree distributions but
are characterized by the same number of available
\emph{resources}, that is by the same number of vertices and
edges. For the only case where the resulting network is not fully
connected, we have only considered the giant component.

In Fig.3 the number of delivered packets, or throughput, has been
plotted as a function of the generation rate, $\lambda$, for the
three considered topologies. Scale-free networks show the least
effective performances in that the number of delivered packets is
lower than for random networks, with Poisson degree distribution.
For scale-free networks, those with $\gamma=2$ are still less
effective than those with $\gamma=3$. Though our analysis is
purely qualitative, we would like to point out that the real
Internet has a power-law degree distribution with $\gamma=2.2$
\cite{Fa:Fa99}.

Notice that the differences among the different considered
topologies, increase for higher values of $\lambda$. In particular
random networks seem to behave better than other networks under
high traffic rates. It is worth noting that this is in strong
agreement with results shown in \cite{Gu:Gu03}.

In Fig.4, the delivery time for packets to reach their destination
has been plotted versus the generation rate, $\lambda$. The
results are in accordance with those for throughput: the highest
delivery time have been achieved for random networks, the lowest
for scale-free networks with $\gamma=2$.

The reason for this is that packets that are stored in the
routers' queues without being delivered to their destination,
increase the time needed for other packets to reach their
destination. Moreover scale-free networks show a vanishing value
of the critical load $\lambda$, i.e. the value of $\lambda$ at
which a phase-transition occurs \cite{Oh:Sa}, with respect to
random graphs.

Consequently, although scale-free networks are characterized by a
shorter characteristic-path-length \cite{Co:Ha03}, they show worst
performances in terms of communication. We conjecture this could
be explained in terms of load distribution. In fact, as we have
already observed, that distribution is characterized by (i) a
decreasing mean (ii) an increasing standard deviation, as the
networks becomes scale-free (especially scale-free networks with
low degree distribution exponents). As a result, since the model
is uncontrolled, most packets will go through the bottle-neck
vertices, localizing the jamming in some zone of the network,
further inducing the onset of congestion.

\section{Conclusions}
\label{sec:conclusion}

We have shown how topological transitions in a given network from
random to scale free affect the load distribution on the network
itself. In particular, we characterised such load distribution in
terms of the average load and its standard deviation. We observed
that as the topological transition takes place, the network
performance worsens and the load tends to become more localised
(higher standard deviation).

Using a novel LRD traffic generation model, we then characterised
the effects of localisation of the load in terms of the typical
parameters used to measure performance of traffic on the network;
namely the number of delivered packets and the average delivery
time. 

\bibliography{biblioridond}
\end{document}